\documentclass{article}
\usepackage{epsfig}
\begin{document}
\renewcommand{\thefootnote}{\fnsymbol{footnote}}
\sloppy
\newcommand{\rp}{\right)}
\newcommand{\lp}{\left(}
\newcommand \be  {\begin{equation}}
\newcommand \bea {\begin{eqnarray}}
\newcommand \ee  {\end{equation}}
\newcommand \eea {\end{eqnarray}}

\title{Response time of internauts}

\author{Anders Johansen \\
CATS, Niels Bohr Institute, Blegdamsvej 17 \\
DK-2100 Kbh. \O, DENMARK \\ 
e-mail: johansen@nbi.dk}

\date{\today}
\maketitle

\begin{abstract}

A new experiment measuring the dynamical response of the Internet population 
to a ``point-like'' perturbation has been performed. The nature of the 
perturbation was that of an announcement, specifically a web-interview on 
   stock market crashes, which contained the URL to the author's articles on 
the subject. It was established that the download rate obeys the relation 
$\approx 1/t$ in qualitative agreement with previously reported results.

\end{abstract}

\section{Introduction}

There can be little doubt that the World-Wide-Web (WWW) provides one of
the most efficient methods for retrieving and distributing information.
As such, it is believed to carry an enormous potential with respect to 
sale and marketing of all kinds of products and is thus of great economic 
interest to most companies. 

A very different aspect of the WWW is that it constitutes a quite unique 
example of a fast evolving social ecology which can be studied in real time. 
There are many examples of self-organising systems in society and Nature, 
where ``less intelligent'' individuals interact creating a ``more intelligent''
whole. Examples are ant hills, flashing colonies of fire-flies, fish schools, 
the stock market, cities, ... A rather troublesome feature with all these 
systems is that they are in general difficult to probe. This is not so with 
the WWW. The fact that it is computer-based and access is unrestricted 
provides a rather unique opportunity to study in real time a fast evolving 
ecology of heterogeneous intelligent agents.

Most studies of the WWW have until now focused on the more easily accessed 
statistical properties, such as the connectivity of the WWW, {\it e.g.},  the 
distributions of outgoing and incoming links \cite{wwwrefs}. However, little 
is known about the internaut population's response to some external or internal
event. With respect to commercial exploitation of the WWW, as well as other 
applications, it is obviously the response of the internaut population to 
some new piece of information which is of interest and not the topology or 
connectivity of the underlying network itself. A rather unique experiment has 
previously been reported in this journal \cite{jpwww}, which studies exactly 
this response. The purpose of the present paper to present the results of a 
similar experiment thus increasing the evidence for a self-similar relaxation
rate \cite{jpwww}. 

The paper proceeds as follows. In the next section, we briefly review the 
previous experiment. In section \ref{uncutexp}, the nature of the present 
experiment is presented and the results compared with those of the previous 
experiment. In section \ref{expla}, a physical setting for the observations 
is presented. Last section concludes.

\section{A previous experiment} \label{jpexp}

An interview with the author and co-worker Didier Sornette on a subject of
broad public interest, namely stock market crashes, was published on the 14th 
of April 1999 in the leading Danish newspaper JyllandsPosten \cite{jpurl}.
The text included the URL's to the author's papers \cite{andersurl} and was 
published both in the paper version and the electronic version on the WWW of 
the newspaper, the latter restricted to subscribers. The URL to the search 
engine of the Los Alamos preprint server, which also contains the author's 
paper on the subject, was included as well. 
 
\begin{figure}[b]
\begin{center}

\epsfig{file=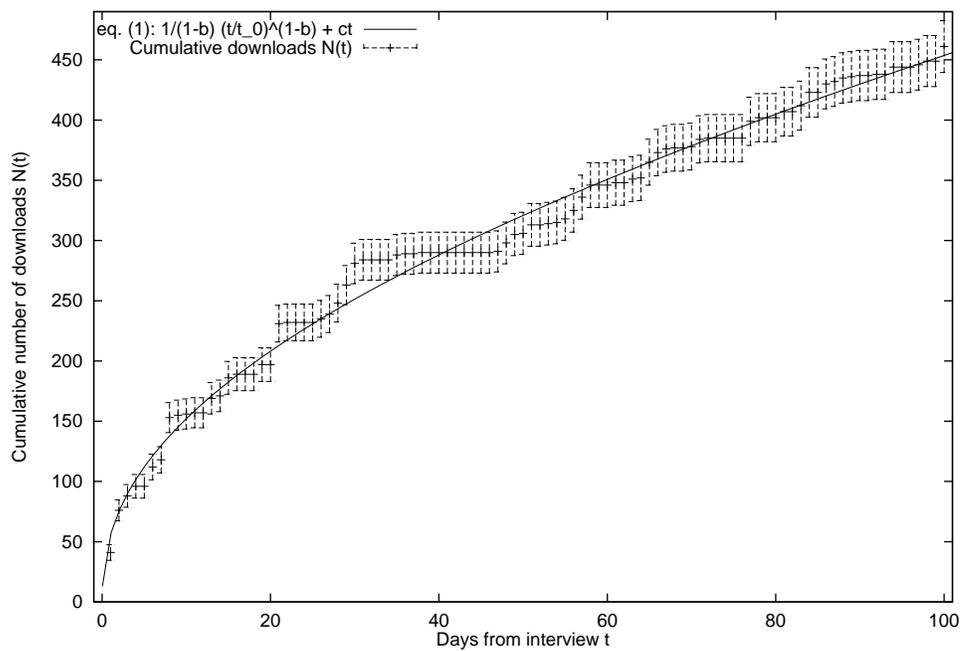}
\caption{\protect\label{jpfig} Cumulative number of downloads $N$ as a
function of time $t$ from the appearance of the newspaper interview on 
Wednesday the 14 April 1999. The fit is $N(t) = \frac{a}{1-b} t^{1-b} + ct$ 
with $b \approx 0.58$.}

\end{center}
\end{figure}

Subsequently, the cumulative number of downloads $N\lp t \rp$ of papers from
\cite{andersurl} was recorded, see figure \ref{jpfig}. It was found that 
$N\lp t \rp$ obeyed a relation
\bea   \label{jpfit}
N\lp t\rp = \frac{1}{1-b} \lp t/t_0 \rp ^{1-b} + ct ,
\eea
over a time period of one hundred days corresponding to a download rate
\bea \label{jprate}
\frac{dN\lp t\rp}{dt} = \frac{t^{b-1}_0}{t^b} + c .
\eea
The last term accounts for a constant background taking into account features
unrelated to the perturbation caused by the interview. From a fit with eq.
(\ref{jpfit}) to the data, see figure \ref{jpfig}, it was found that $b\approx 
0.6$, $t_0 \approx 0.8$ minutes and $c\approx 0.8$ days$^{-1}$.

\section{The present experiment} \label{uncutexp}

The Nasdaq crash culminating on Friday the 14th of April 2000 caught many 
people with surprise and shook the stock market quite forcefully.  As always, 
many different reasons for the crash were given ranging from the anti-trust 
case against MicroSoft to an ``irrational exuberance'' of the participants
on the stock market. The author and co-worker Didier Sornette had also bid 
for the cause \cite{nascrash}, which was made public on Monday the 17th of 
April 2000 on the Los Alamos preprint server. As a result, a forty minute 
interview with the author called ``The World (Not) According to GARCH'' was 
published on Friday the 26th of May 2000 on a ``radio website'' \cite{uncut}. 
As in the experiment presented in section \ref{jpexp}, the URL to the author's 
papers \cite{andersurl} was announced making it clear that work on stock market
crashes in general and the recent Nasdaq crash in particular could be found 
using the posted URL. The URL of the Los Alamos preprint server was not 
included on this occasion. As in the previous experiment, the the number of 
downloads of papers as a function of time from the appearance of the interview 
was recorded.

A few {\it a priori} important differences between the two experiments should 
be held forward. The web-interview appeared on a {\it single} web-site and only
a {\it single} source was provided for the download of the author's papers. In 
the experiment described in the previous section, the interview was published 
in the electronic as well as the paper version of the newspaper JyllandsPosten.
Hence, it was initially not freely available, since you either had to buy the 
newspaper or be a subscriber. A stronger response to the web-interview should 
be expected due to an audience which is much more oriented towards the subject 
than the average Danish newspaper reader. Furthermore, since the web-interview 
was in English it had potentially a larger audience than its Danish 
counter-part.

\begin{figure}
\begin{center}

\epsfig{file=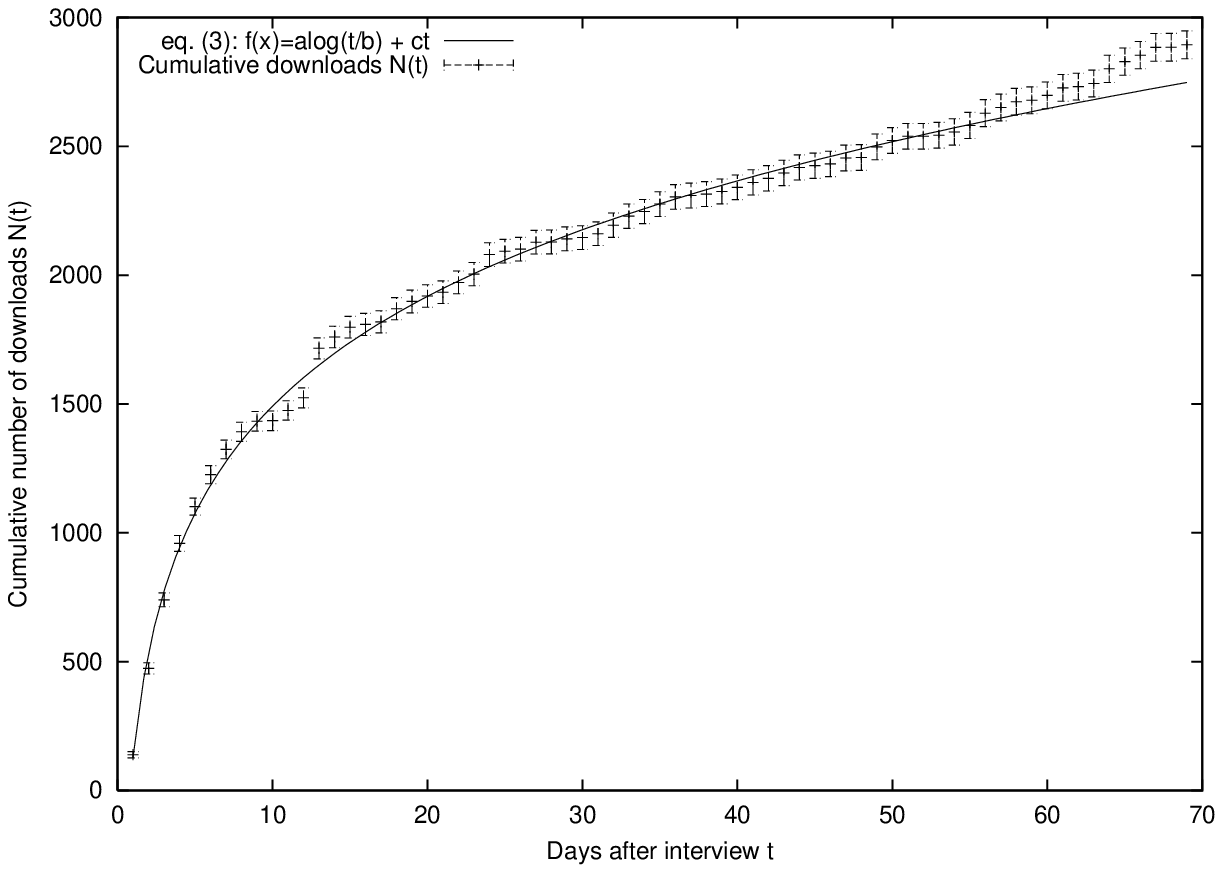}
\caption{\protect\label{uncutfig1} Cumulative number of downloads $N\lp t\rp$ 
as a function of time $t$ from the appearance of the Web-interview on Friday
the 26 May 2000. The fit is $N\lp t\rp = a\ln\lp t/\tau \rp + ct$.}

\vspace{5mm}

\epsfig{file=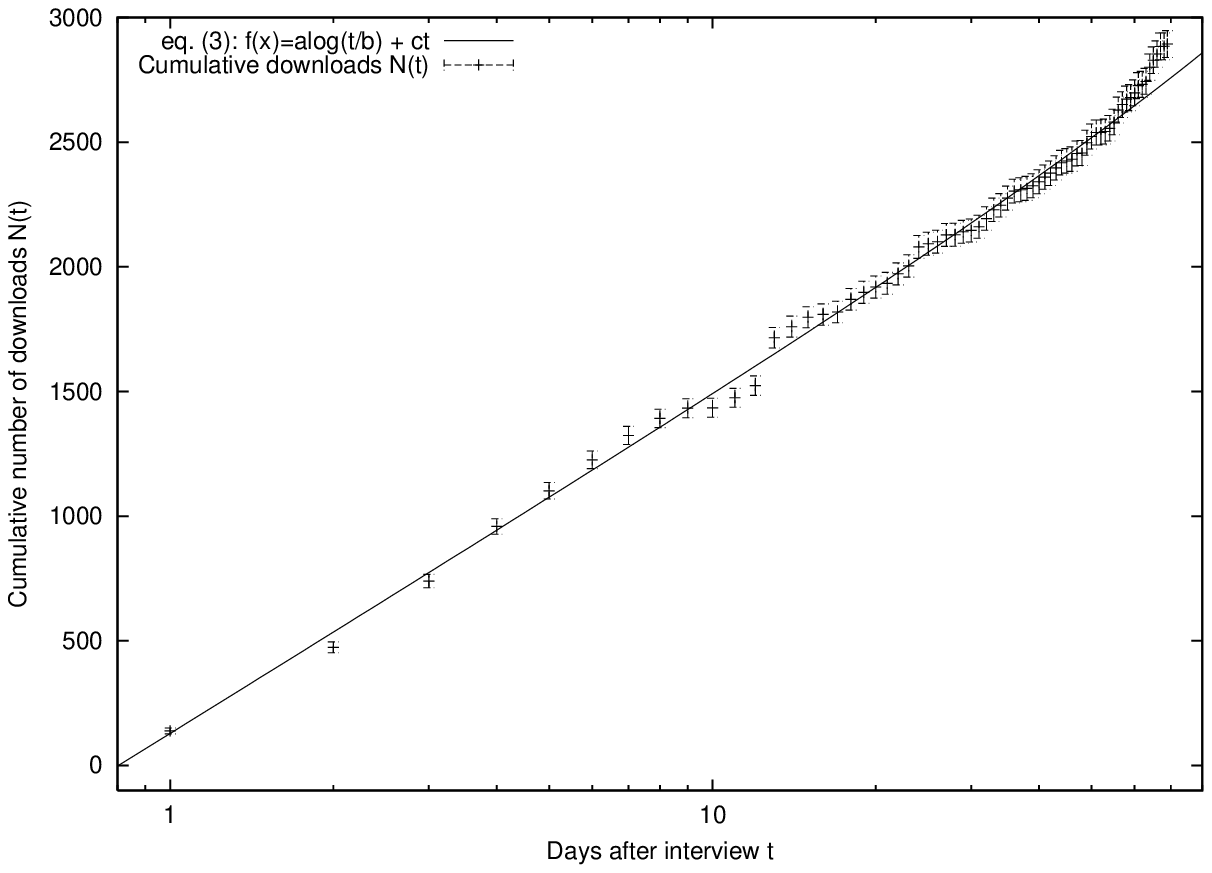}
\caption{\protect\label{uncutfig2} Cumulative number of downloads $N\lp t\rp$ 
as a function of time $t$ from the appearance of the Web-interview on Friday
the 26 May 2000. The fit is $N\lp t\rp = a\ln\lp t/\tau \rp + ct$.}

\end{center}
\end{figure}

In figures \ref{uncutfig1} and \ref{uncutfig2}, the cumulative number of 
downloads $N\lp t\rp$  as a function of time $t$ after the appearance of the 
interview are shown on a linear and semi-logarithmic scale. The error-bars 
are simply taken as the square-root of $N\lp t\rp$. The data is over $\approx
60$ days surprisingly well-captured by the relation
\be \label{downeq}
N\lp t\rp = a\ln\lp t/t_0 \rp + ct~ ,
\ee
with $a \approx 583$, $t_0 \approx 0.80$ days and $c \approx 2.2$ days$^{-1}$,
corresponding to a download rate 
\be \label{uncutrate}
\frac{dN\lp t\rp}{dt} = \frac{a}{t} + c .
\ee
Thus, we again obtain a power law relaxation of the download rate, however, 
with a different exponent. 

After $\approx 60$ days, we see how the data breaks away from the fitted line. 
The reason is the appearance of \cite{andersurl} on the Social Science Research
Network server (http://WWW.SSRN.COM) causing a second perturbation\footnote{
That a second advertisement of \cite{andersurl} had appeared on SSRN became 
quite evident when the author started receiving e-mail requests for preprints 
citing SSRN as information source.}. Hence, the experiment became influenced 
by a second perturbation of non-negligible impact and was consequently halted 
after 69 days.

\section{Interpretation of results} \label{expla}

The results shown in figures \ref{jpfig}, \ref{uncutfig1} and \ref{uncutfig2}  
suggests that the rate of downloads is that of a power law with a constant 
background
\bea
\frac{dN\lp t\rp}{dt} = \frac{a}{t^b} + c .
\eea
where $b$ is not universal\footnote{In the present context, $b$ varies with 
{\it e.g.}, the subject and the language of the interview.} Such power law 
dependence of the rate of ``events'' are found in other areas of Physics. 
Two well-known examples are Omori's law for the rate of aftershocks as a 
function of time elapsed since the main event \cite{omori} and relaxation of 
spin glasses subjected to a magnetic field \cite{trap}. 

At time $t$ after the appearance of the interview, the internaut community 
consists of two populations, namely those who have not downloaded a paper from 
\cite{andersurl} and those who have. The transition from the first state to 
the second demands the crossing of some threshold specific to each internaut. 
We thus imagine that the announcement of the URL plays the role a ``field'' to 
which the internaut population is subjected and study the relaxation process 
by monitoring the number of downloads as a function of time since the field 
was applied. Within this frame-work \cite{jpwww}, we may view the process as 
a diffusion process in a random potential, where the act of downloading is 
similar to that of a barrier-crossing in the Trap model \cite{trap} of spin 
glasses. In this model, the distribution of trapping times $\tau_i$ is a power 
law 
\be  \label{trapeq1}
P\lp \tau \rp \propto \tau^{-\lp 1+x\rp }, 
\ee
where $x$ depends of the experimental conditions, {\it e.g.}, $x=T/T_g$ for a 
spin glass where $T_g$ is the temperature of the glass transition. For 
$ x < 1$, the mean trapping time $\langle \tau \rangle\equiv \int_0^\infty
\tau P\lp \tau \rp d\tau$ diverges and, as a consequence, the time needed 
for the system to relax is infinite. This is known as ``weak breaking of 
ergodicity'' in terms of spin glasses and is the hallmark of ``aging'' in
anomalous relaxation processes.

For finite times we may proceed as follow. The parameter $x$ is determined by
$$
\langle \tau \rangle = \int^{\tau_{max}}_{t_{min}}  
{\tau \over \tau^{1+x}} d\tau \sim \Delta t_{max}^{1-x},~~x<1 .
$$
The maximum $\tau_{max}$ is typically given by \cite{trap} 
$$
N \int_{\tau_{max}}^{\infty} {d\tau \over \tau^{1+x}} \sim 1~,  
$$
where N is the number of barrier crossings, {\it i.e.}, the number of 
downloads. Hence,
$$
\tau_{max} \sim N^{1 \over x}~, 0 < x < 1 
$$
Thus $t = N \langle \Delta t \rangle \sim N^{1 \over x}$, which finally gives
\be \label{trapeq2}
N \sim t^x \Rightarrow \frac{dN\lp t\rp}{dt} \sim t^{x-1}
\ee
In the case of the second experiment, the logarithmic growth of $N(t)$ formally
corresponds to the case $x=0$.  However, in the present framework this case 
correspond to $x \ll 1$, which gives  the same behaviour up to first order. 
Comparing relation (\ref{trapeq2}) with (\ref{jprate}) and (\ref{uncutrate}), 
we find $1-b = x < 1$ in both experiments. This means that the longer since the
last download, the longer the expected time till the next one \cite{sorknop}. 
In other words, any expectation of a download that is estimated today depends 
on the past in a manner which does not decay. That the relaxation exponent $b$ 
is less than one has an important consequence, namely non-stationarity and 
``aging'' in the technical sense of a breaking of ergodicity \cite{trap}. 

Within the proposed framework, one may speculate on the nature of the ``field''
created by the interview and its effect on the internaut population. Obviously,
the act of downloading is a consequence of a) that the information of the URL 
is available to the internaut and b) that the knowledge of the URL results in 
a download by the internaut. The first contribution relates to how information 
spreads through the internaut population, the second to a ``response-time'' 
specific to each individual internaut. However, these two contributions are  
consequences of the same underlying processes, since the spread of information 
on the WWW is a consequence of the individual internaut's decision to post a 
link, send e-mails, ... which again is governed by the internauts' 
response-times. In fact, no download would take place if

\begin{enumerate}

\item  nobody decided to visited the site where the URL was posted.

\item  nobody decided to read/listen to the interviews.

\item  nobody decided to use the URL after understanding the subject.

\item  nobody decided to download after using the URL.

\item  nobody decided to to tell others about the interview and URL.

\item  nobody decided to use the URL after being told about it and the 
       interview. 

\item  ...

\end{enumerate}

These considerations strongly suggest that eq. (\ref{trapeq1}), and 
consequently eq. (\ref{trapeq2}), reflects the variation in response-time 
of internauts when confronted with a new piece of information. Furthermore, 
if the contributions of 5) and 6) are negligible this suggest that in the 
absence of deadlines internauts (humans) have a power law distribution of 
waiting times before deciding to act on a new piece of information. Each 
individual thus has a ``psychological barrier'' which must be crossed in 
order for the individual to act and the time needed to cross this barrier 
is power law distributed throughout the internaut (human) population.

\section{Concluding remarks}

The two rather unique experiments presented here in both cases shows a 
self-similar dynamical response of the internaut population to a 
perturbation, even though {\it a priori} the nature of the perturbations 
were quite different. Specifically, we have presented two experiments where 
the rate of downloads $N\lp t\rp$ from a site obeys a power law
$$
\frac{dN\lp t\rp}{dt} \sim  t^{-b}. 
$$
as a function of time $t$ after an advertisement on the WWW of that site's 
URL. 

In Physics, the fundamental approach for studying a given system is to 
introduce a well-controlled perturbation on the system and then study 
{\it e.g.}, its response function, its susceptibility and its relaxation 
dynamics. In the social fields, it is generally impossible to control the 
effective perturbation, since it has its origin in a variety of complex 
sources with unknown internal interactions as well as with the system itself.
It is therefore very difficult to study the response of a social system in  
a controlled manner. Here, we have discussed two quite unique cases of a 
perturbation of a social self-organising system, which is controlled enough 
to allow the use of a Physics methodology. It uses a point-like perturbation 
and then measures the relaxation of the system as a function of time by 
monitoring a single site. This as well as the fact that the announcements 
(perturbations) were focused on a quite visible problem (financial crashes) of 
interest to a broad class of people produces a large impact compared to the 
background and makes the two experiments very ``clean''.

In summary, we have reported two unique experiments in real time where
the dynamical relaxation of the internaut population on the WWW in response 
to a point-like perturbation exhibits a slow power law decay similar to what 
have been observed for earthquake aftershocks and relaxation of spin glasses 
in the glass phase. In both experiments, a power law relaxation was obtained
thus quantifying the internaut population response to a ``perturbation'', 
{\it i.e.}, the impact of an advertisement on the internaut population. 
With respect to applications, it is obviously the impact of some new piece
of information on the internaut population which is of interest and not the 
topology and connectivity of the network as such. By heuristic arguments, 
it was argued that the power law response of the internaut population is a 
fingerprint of the psychological processes initiated by the reception of 
new information and ending with the triggering of some action based on this 
new information. Specifically, the two experiments suggest that in the absence
of deadlines the response-time of internauts (humans) to new information have 
a power law distribution throughout the population. The existence of such a 
fat-tailed distribution of response-times of ``agents'' has quite important 
implications for the modelling of social systems such as the financial markets.
\\

{\bf Acknowledgement:} The author thanks Didier Sornette for stimulating
discussions.

\end{document}